# Improved Ionic Model of Liquid Uranium Dioxide

Victor Gryaznov[(1)], *Igor Iosilevski[(2)], Eugene Yakub[(3)], Vladimir Fortov[(4)],
Gerard J. Hyland[(5)] and Claudio Ronchi[(6)]

[1] Institute of Chemical Physics of RAS, 142432, Chernogolovka, Russia
[2] Moscow Institute of Physics and Technology (State University), 141700, Moscow Russia
[3] Odessa State Medical University, Odessa, Ukraine
[4] High Energy-Density Research Centre of RAS, 127412, Moscow, Russia
[5] University of Warwick, Coventry, CV4 7AL, U.K.
[6] European Commission, Joint Research Centre, Institute for Transuranium, Karlsruhe, Germany,

**Abstract.** The paper presents a model for liquid uranium dioxide, obtained by improving a simplified ionic model, previously adopted to describe the equation of state of this substance [1]. A "chemical picture" is used for liquid $UO_2$ of stoichiometric and non-stoichiometric composition. Several ionic species are considered here: $U^{5+}$, $U^{4+}$, $U^{3+}$, $O^{2-}$ and $O^{-}$. The ions are described as charged hard-spheres of different diameters. Coulomb interaction of ions is taken into account according to the modified Mean Sphere Approximation (MSA). The main result of the new model is the appearance of natural "plasma" equivalent, which, from the theory, is directly related to the definition of oxygen potential in liquid $UO_{2\pm x}$. The features of the model make it possible to describe *non-congruent* phase equilibrium (and evaporation) in uranium dioxide, as well as other relevant phenomena characterising the phase equilibrium in chemically active matter. First calculation results are discussed.

## 1. INTRODUCTION

The equation of state (EOS) of gaseous and liquid uranium dioxide is of primary importance for nuclear safety calculations [2-9]. The primary objective of this work is to predict the equilibrium pressure and the composition of the equilibrium vapours over boiling $UO_{2\pm x}$. *Non-congruence*, i.e. coexistence of phases having different O/U stoichiometry, represents one of the observed remarkable features of the evaporation of uranium dioxide. This property is closely related to the *oxygen chemical potential*, which is a fundamental quantity in the thermodynamic behaviour of the solid phase of $UO_{2\pm x}$ [7]. A lot of attempts have been made during the last decades to describe theoretically the phase coexistence in uranium dioxide. For example, Mistura, Magill and Ohse [8], postulating isomorphism between liquid $UO_2$ and $CO_2$, started from the Law of Correspondent States to construct their EOS. Another popular phenomenological approach consists in separating the description of the vapour from that of the condensed phases (see Green and Leibowitz [6]); the Gibbs free energy of liquid $UO_2$ is estimated from its heat capacity, $C_p(T)$, and is used along with an oxygen potential model extrapolated from a model of solid uranium dioxide (see Hyland [7]). So far, the most successful attempt to describe non-congruent evaporation in $UO_2$ is the work of Fischer [9], who used the Significant Structures Theory (SST) developed by Eyring [10], in combination with an independent model of oxygen potential in the liquid.

The present work, performed in the frame of an international INTAS Project [2], is aimed at constructing an equation of state of uranium dioxide, intended to be valid for both the vapour and the liquid phase, be based on a realistic theoretical approach, and be eventually validated by satisfactory agreement with the complete set of updated experimental data.

## 2. IONIC MODEL

Uranium dioxide in the crystalline phase exhibits several features typical for substances with predominantly ionic bonding. Consequently, it is presumed to maintain this property also on melting (Bhuiyan and March, Sindringre and Gillan [11, 12]). For instance, Sindringre and Gillan [12] have successfully carried out molecular dynamics (MD) computer simulations of $UO_2$ using non-empirical inter-ionic interaction potentials both in the solid and liquid state.

*Corresponding author: ilios@orc.ru

Different ways were initially attempted [1,2] in the reported project, since a wide front of approach was actually justified by the demonstrated formal equivalence of the representation of *quasi-molecular* [2-4] and *pure ionic* [1,2] atomic interactions in the condensed phase. For two of the simplified ionic models examined, calculations were carried out earlier [1]. Both models were restricted by the assumption of *invariant stoichiometry* of $UO_2$; the liquid was hence depicted as an electrically neutral mixture of two charged species, $U^{4+}$ and $O^{2-}$, both having the same hard-sphere diameter, $\sigma$. Within the *Restricted Primitive Ionic Model* (RPIM) only hard-sphere and Coulomb interactions of ions were considered. Additional soft short-ranged attractive interaction of van-der-Waals' type was taken into account in the *Improved Restricted Ionic Model* (IRIM).

## 2.1. Equilibrium Composition of Liquid and Vapour

The present, further improved, ionic model was extended to describe non stoichiometric equilibrium compositions of liquid $UO_{2\pm x}$, by taking into account possible formation of additional ionic species: two uranium ions, $U^{5+}$ and $U^{3+}$, and one oxygen ion, $O^-$. This extension is equivalent to assuming possible *electronic exchanges* between all ions present in the mixture, for example, through the reactions: $2U^{4+} = U^{5+} + U^{3+}$, $O^{2-} + U^{4+} = O^- + U^{3+}$, etc. The equilibrium concentrations of all the considered ions ($U^{3+}$, $U^{4+}$, $U^{5+}$, $O^{2-}$ and $O^{1-}$) in liquid $UO_{2\pm x}$ (of arbitrary stoichiometry) were calculated with the well-known formalism of ionisation equilibrium ("Chemical Picture"), using *non-ideal* EOS and Saha equations (see, e.g., [13]). The same approach was applied to describe the chemical and ionisation equilibrium in the vapour phase. The composition of the equilibrium vapour over $UO_{2\pm x}$ includes molecules: $UO$, $UO_2$, $UO_3$, $O$ and $O_2$, as well as molecular ions $UO_2^+$ and $UO_3^-$. Therefore, natural "plasma" equivalent appears in this theory of liquid $UO_{2+x}$, reflecting an implicit "oxygen potential" model. Furthermore, a description of *non-congruent* phase equilibrium (evaporation) is obtained:

$$(\mu[O])_{vapor} = (\mu[O])_{liquid} = \{\mu[O^{2-}] - 2\mu[e^-] = \mu[O^{2-}] + \mu[U^{5+}] - \mu[U^{3+}]\}_{liquid} \quad (1)$$

$$(\mu[UO_2])_{vapor} = (\mu[UO_2])_{liquid} = \{2\mu[O^{2-}] + \mu[U^{4+}]\}_{liquid} \quad (2)$$

Consistent calculations of chemical, ionisation-, and phase-equilibrium were performed with a computer code of the "SAHA" family [13], modified for this special purpose. Thermochemical data and excitation partition functions from the IVTAN-Database [14, 2] have been used as input for all the atomic, molecular and ionic species.

It could be proved that the model, corresponding to a *highly ionic* representation of the equilibrium composition of liquid $UO_{2\pm x}$, could be successfully applied to describe the *non-congruent* phase equilibrium. Finally, the model results are compatible with a *quasi-molecular* representation of the vapour composition over boiling uranium dioxide.

## 2.2. Interaction (Non-Ideality) Corrections in Ionic Models

Ionic interactions were assumed to be pair–additive, and consisting of two parts: short–range hard-sphere repulsion, and Coulomb long-range interaction. A non-ideality correction due to short-range inter-ionic and inter-molecular repulsion was taken into account using Mansoori's approximation for Hard Spheres Mixture [15]. One-parametric form of the Coulomb correction was also applied; this consists of a superposition of MSA (Mean Spherical Approximation) + DHSA (Debye-Hückel for Charged Sphere Approximation) [16, 17] combined with "one-fluid" approximation for the case of an ionic mixture with different diameters.

$$\frac{\Delta F_C}{RT} \equiv \left(\frac{\Delta F_C}{RT}\right)_{DHLL} \theta(x,\nu) \cong \left(\frac{\Delta F_C}{RT}\right)_{DHLL} \theta(x); \qquad x \equiv <\sigma>/r_D \qquad r_D^2 \equiv \frac{4\pi e^2}{VkT}\sum_{\alpha=1}^{2} N_\alpha Z_\alpha^2 \quad (3)$$

$$\theta(x) = (3\alpha/x^3)[\text{Ln}(1 + x) - x + x^2/2] + (1 - \alpha)\{[2(1 + 2x)^{3/2} - 3x^2 - 6x - 2]/x^3\} \quad (4)$$

$$x \equiv \langle\sigma\rangle/R_D \qquad \langle\sigma\rangle \equiv [(\sum n_i \sigma_i^3)/(\sum n_i)]^{1/3} \quad (5)$$

Here $(\Delta F_C/VkT)_{DHLL} \equiv -(12\pi r_D^3)^{-1}$ is the Debye–Hückel Limiting Law.

## 2.3. Calibration of the model

From the known values of density, $\rho_m$, entropy, $S_m$, Gibbs free energy, $G_m$, and oxygen potential, $\mu_O$ [7], of liquid stoichiometric $UO_{2.00}$ at the melting temperature, $T = 3120$ K, the following parameters were fitted:
  – Reference ionic diameters, $\{\sigma^{(0)}\}$ $\qquad$ $[\sigma(U^{4+}) = \sigma(O^{2-}) = \sigma^{(0)}]$
  – "Strength" of Coulomb correction at $x \gg 1$ (parameter $\alpha$ in approximation (3,4)
  – Dispersion of ionic diameters $\{\sigma_i/\sigma^{(0)}\}$
  – Energy of creation of ion $O^{2-}$ (in fact, this value is uncertain).

The values of the fitted parameters and a comparison of the predicted thermodynamic properties for liquid and vapour with existing experimental data are presented in Tables 1 and 2.

**Table 1.** Comparison of predictions of present ionic model with experimental data [18], results of MD numerical simulation [12] and values recommended by INSC [19] and IVTAN [2,14] databases.

| $T$, K | 3120 (exp.) | 3120 (MD) | | 3120 | 4000 | 5000 | 6000 |
|---|---|---|---|---|---|---|---|
| $\rho_{liq}$, g/cm³ | 8,87 | 7.20 | Ionic model [a] *INSC Database* | 8.88 [b] *8.86* | 7.99 *8.04* | 7.16 *7.11* | 6.42 *6.19* |
| $S_{liq}$, J/g K | 1.183 [c] | – | Ionic model | 1.183 [b] | 1.285 | 1.381 | 1.466 |
| - $G_{liq}$, kJ/g | 10.03 [c] | – | Ionic model | 10.03 [b] | 11.121 | 12.456 | 13.879 |
| $C_P$, J/kgK | 440 | 360 | Ionic model *Experiment [18]* | 410 *450 ± 60* | 416 *320 ± 70* | 446 *320 ± 80* | 486 *360 ± 90* |
| $\alpha_P \cdot 10^4$, K⁻¹ | 1.05 | 0.2 | Ionic model *INSC Database* | 1.27 *1.05* | 1.13 *1.15* | 1.09 *1.30* | 1.11 *1.50* |
| $\beta_T \cdot 10^5$, (MPa)⁻¹ | 4.10 | 8.13 | Ionic model | 8.13 | 8.82 | 9.81 | 11.2 |
| **Liquid composition** | | | | | | | |
| $[n(U^{3+})/n(U^{4+})]$ | – | 0 | Ionic model | 0.00129 | 0.00920 | 0.0378 | 0.102 |
| $[n(O^-)/n(O^{2-})]$ | – | 0 | Ionic model | 0.00069 | 0.00460 | 0.0191 | 0.050 |
| **Vapor composition**[d] | | | | | | | |
| Total (O/U) ratio | – | – | Ionic model | 2.40 | 2.43 | 2.71 | 3.65 |
| Total vapor pressure kPa | 3.90 ±50% | – | Ionic model *INSC Database* | 3.81 *4.69* | 124 *189* | 1580 *2280* | 10420 *10910* |

[a] Reference ionic diameters $\sigma^{(0)} = 1.99$, $10^{-10}$ m:
[b] Fitted from experimental and tabulated $\rho_o$, $S_o$ and $G_o$ values at $T = 3120$ K.
[c] Tabulated values from IVTAN database [2].
[d] Boiling conditions $\{(O/U)_{Liquid} = 2.0; (O/U)_{Vapor} \neq 2.0\}$

**Table 2.** Dispersion of the diameters of the ionic species in the ionic model for liquid $UO_{2\pm x}$
(The diameter of $U^{4+}$, $\sigma^{(0)} = 1.99, 10^{-10}m$, is taken as reference)

| Ion | $U^{3+}$ | $U^{4+}$ | $U^{5+}$ | $U^{6+}$ | $O^{2-}$ | $O^{1-}$ | |
|---|---|---|---|---|---|---|---|
| $\sigma_i/\sigma(U^{4+})$ | 1.13 | 1.00 | 0.87 | 0.84 | – | – | [20][*] |
| $\sigma_i/\sigma^{(0)}$ | 1.11 | 1.00 | 0.88 | – | 1.00 | 0.91 | Ionic Model |

[*] $R[U^{3+} //U^{4+} //U^{5+} //U^{6+}] = [0.1165 //0.103 //0.090 //0.087]$ nm.

## CONCLUSIONS

– Liquid uranium dioxide ($UO_{2+x}$) is an interesting example – and typical for its class of compounds - of a *chemically active*, strongly coupled Coulomb system. Ionic models, previously developed to describe the thermodynamic properties of liquid $UO_2$ at invariant composition, have been extended to include species other than the two basic ions, $U^{4+}$ and $O^{2-}$. Liquid $UO_{2\pm x}$ was considered as an equilibrium multi-component and highly ionised mixture: $\{U^{3+} + U^{4+} + U^{5+} + O^{2-} + O^-\}$. As a consequence of this extension, the analogue of the "*oxygen potential*" of liquid $UO_{2\pm x}$ is naturally introduced. This clears the way to the description of *non-congruent* evaporation, an essential topic in the phase equilibrium of chemically active systems.

– The proposed model successfully describes selected experimental thermodynamic parameters of $UO_2$ at the melting temperature. Furthermore, the model provides a satisfactory qualitative description of the properties of $UO_2$ extrapolated to very high temperatures, and, furthermore, correctly predicts the decrease of the ionisation degree with increasing temperature, in accordance with the results obtained from numerical simulations of simpler ionic systems (e.g., $Na^+ + Cl^-$ [21]).

– From the assumption of the ionic character of liquid $UO_2$ it follows that, starting from a cold and dense mixture of *highly ionised* uranium and oxygen - and going along the coexistence curve - the state of the system should be continuously transformed to match that of the cold *neutral* vapour. The features and location of this transformation in uranium dioxide is, in this model context, one of the most important unsolved problems yet. Efforts in this direction are continuing.

### Acknowledgements

The work was supported by Grant INTAS 93-0066 and Grant 2550. "Universities of Russia"